\documentclass[onecolumn,showpacs,amsfonts,aps,prc,nofootinbib,floatfix,%
superscriptaddress]{revtex4}

\usepackage{float}

\usepackage{amsmath}
\usepackage{bm}
\usepackage{graphicx}

\usepackage{epsfig}
\newcommand{\beq}{\begin{equation}}
\newcommand{\eeq}{\end{equation}}
\newcommand{\bea}{\vspace{0.25cm}\begin{eqnarray}}
\newcommand{\eea}{\end{eqnarray}}

\newcommand{\ro}{\mbox{{\boldmath
$\rho$}}}

\newcommand{\qb}{\mbox{{\bf
q}}}
\newcommand{\pb}{{{\bf p}}}

\newcommand{\bb}{{{\bf b}}}

\def\lsim{\mathrel{\rlap{\lower4pt\hbox{\hskip1pt$\sim$}}
    \raise1pt\hbox{$<$}}}         
\def\gsim{\mathrel{\rlap{\lower4pt\hbox{\hskip1pt$\sim$}}
    \raise1pt\hbox{$>$}}}         

\begin{document}


\title{Jet quenching with $T$-dependent running coupling
}
\date{\today}

\author{B.G. Zakharov}

\address{
L.D.~Landau Institute for Theoretical Physics,
        GSP-1, 117940,\\ Kosygina Str. 2, 117334 Moscow, Russia
}
\address{Steklov Mathematical Institute, Russian Academy of Sciences, Gubkin str. 8, 119991
Moscow, Russia}

\begin{abstract}
  We perform an analysis of jet quenching in heavy ion collisions
  at RHIC and LHC energies with the temperature dependent running
  QCD coupling. Our results show that the $T$-dependent QCD coupling
  largely eliminates the difference between the optimal
  values of $\alpha_s$ for the RHIC and LHC energies.
It may be viewed as direct evidence of the increase
of the thermal suppression
of $\alpha_s$ with rising temperature.

\end{abstract}
%

\maketitle

{\bf 1. Introduction}.
It is accepted that the strong suppression of the high-$p_T$
particle spectra in $AA$ collisions (usually called the jet quenching)
observed at RHIC and LHC, is due to parton energy loss
(radiative and collisional) in the quark-gluon plasma (QGP).
The jet quenching is one of the major signals of the QGP
formation in relativistic $AA$ collisions.   
The main contribution to the parton energy loss
comes from the radiative mechanism due to induced gluon emission
\cite{BDMPS1,LCPI1,W1,GLV1,AMY1}. 
The effect of the collisional energy loss
turns out to be relatively weak \cite{BSZ,Z_coll}.

The available pQCD approaches to the radiative energy loss
\cite{BDMPS1,LCPI1,W1,GLV1,AMY1}
are limited to the one gluon emission.   
The effect of multiple gluon radiation is usually accounted for
in the approximation of independent gluon emission \cite{RAA_BDMS}.
Altogether, the pQCD calculations within this approximation
give a rather good agreement with the jet quenching
data from RHIC and LHC (see e.g. \cite{RAA20} and references therein).
However, it was found that, in the formulation with a unique
temperature independent
QCD coupling, the simultaneous description
of the RHIC and LHC data requires to use somewhat smaller $\alpha_s$ at
the LHC energies \cite{RAA11JP,RAA11,RAA13,RAA20}
(in \cite{JETC_qhat,Salgado_qhat}
a similar difference between jet quenching at RHIC and LHC energies,
has been found in terms of the transport coefficient $\hat{q}$).
In \cite{RAA11JP,RAA11,RAA13,RAA20} this fact has been demonstrated
within the light-cone path integral (LCPI) approach to induced
gluon emission \cite{LCPI1},
using the method
developed in \cite{RAA04,RAA08},
for running $\alpha_s$ which is frozen
at low momenta at some value $\alpha_{s}^{fr}$.
There it was found that the RHIC data support
a significantly larger value of
  $\alpha_{s}^{fr}$ than the LHC data.
One of the reasons for this difference
may be somewhat stronger thermal suppression
of the effective QCD coupling in a hotter QGP at the LHC energies.
To draw a firm conclusion on this possibility
it is highly desirable to perform calculations with a temperature
dependent $\alpha_s$. And of course, it is clear that, even without respect
to the problem with a joint description of the RHIC and LHC jet quenching
data, an observation of the temperature dependence
of $\alpha_s$ from  the jet quenching data
would be of great importance on its own.
The case of the $T$-dependent coupling has not been
discussed so far in the literature on jet quenching.
The purpose of this work is to perform such an analysis.
We adapt the LCPI formalism to the case of the $T$-dependent
running $\alpha_s$, and perform a joint analysis of
the jet quenching data from RHIC on $0.2$ TeV Au+Au collisions
and from the LHC on $5.02$ TeV Pb+Pb collisions.

{\bf 2. Theoretical framework}.
We will consider the central rapidity region around $y=0$.
Our method for calculating the nuclear modification factor $R_{AA}$ is
similar to the one used in our previous jet quenching analyses
\cite{RAA08,RAA13,RAA20}. Therefore, we only outline its main points.
%
We write the nuclear modification factor $R_{AA}$ 
for given impact parameter $b$ of $AA$ collision,
the hadron transverse momentum $\pb_T$
and rapidity $y$
as
\beq
R_{AA}(b,\pb_{T},y)=\frac{{dN(A+A\rightarrow h+X)}/{d\pb_{T}dy}}
{T_{AA}(b){d\sigma(N+N\rightarrow h+X)}/{d\pb_{T}dy}}\,,
\label{eq:10}
\eeq
where $T_{AA}(b)=\int d\ro T_{A}(\ro) T_{A}(\ro-\bb)$,
$T_{A}(\ro)=\int dz \rho_A(\sqrt{\rho^2+z^2})$ is the nuclear thickness function
(with $\rho_A$ the nuclear density).
The nominator on the right hand side of (\ref{eq:10})
is the differential yield of the 
process $A+A\to h+X$ (we omit for clarity the arguments $b$ and $\pb_T$).
It can be written via the medium-modified hard cross section
$d\sigma_{m}/d\pb_{T} dy$ in the form
\beq
\frac{dN(A+A\rightarrow h+X)}{d\pb_{T} dy}=\int d\ro T_{A}(\ro+\bb/2)
T_{A}(\ro-\bb/2)
\frac{d\sigma_{m}(N+N\rightarrow h+X)}{d\pb_{T} dy}\,.\,\,\,
\label{eq:20}
\eeq
We write $d\sigma_{m}/d\pb_{T} dy$ as 
\beq
\frac{d\sigma_{m}(N+N\rightarrow h+X)}{d\pb_{T} dy}=
\sum_{i}\int_{0}^{1} \frac{dz}{z^{2}}
D_{h/i}^{m}(z, Q)
\frac{d\sigma(N+N\rightarrow i+X)}{d\pb_{T}^{i} dy}\,,\,\,\,
\label{eq:30}
\eeq
where $\pb_{T}^{i}=\pb_{T}/z$ is the transverse momentum
of the initial hard parton, 
${d\sigma(N+N\rightarrow i+X)}/{d\pb_{T}^{i} dy}$ is the
ordinary hard cross section, and 
$D_{h/i}^{m}(z,Q)$ is the medium-modified fragmentation function
for transition of a parton $i$ with the virtuality $Q\sim p^i_T$ to the
final hadron $h$. The fragmentation functions $D_{h/i}^{m}(z,Q)$
accumulate the medium effects. They depend crucially
on the QGP fireball density profile along the hard parton trajectory.
We use somewhat improved method of \cite{RAA08} for evaluation
of $D_{h/i}^{m}(z,Q)$ via the one gluon induced spectrum
in the approximation of independent gluon emission \cite{RAA_BDMS}.
We refer the interested reader to \cite{RAA20} for details.
Also there, a detailed description
of the technical aspects of the implementation of formulas
(\ref{eq:10})--(\ref{eq:30})
can be found.

We turn now to the method for incorporation
of the temperature dependent coupling in calculating  the induced gluon
spectrum. Let us consider the case of $q\to g q$ process.
In the LCPI formalism \cite{LCPI1}
the gluon spectrum in $x=E_g/E_q$ for $q\to g q$ process
can be written as \cite{RAA04}
\beq
\frac{d P}{d
x}=
\int\limits_{0}^{L}\! d z\,
n(z)
\frac{d
\sigma_{eff}^{BH}(x,z)}{dx}\,,
\label{eq:40}
\eeq
where $n(z)$ is the medium number density, $d\sigma^{BH}_{eff}/dx$ 
is an effective Bethe-Heitler
cross section accounting for both the Landau-Pomeranchuk-Migdal
and the finite-size effects. Note that for the midrapidity region
$y=0$, the longitudinal coordinate $z$ in (\ref{eq:40}) coincides with
the proper time $\tau$ for evolution of the QGP fireball. 
At fixed coupling $d\sigma^{BH}_{eff}/dx$ can be written as \cite{RAA04}  
\beq
\frac{d
\sigma_{eff}^{BH}(x,z)}{dx}=-\frac{P_{q}^{g}(x)}
{\pi M}\mbox{Im}
\int\limits_{0}^{z} d\xi \alpha_{s}
\left.\frac{\partial }{\partial \rho}
\left(\frac{F(\xi,\rho)}{\sqrt{\rho}}\right)
\right|_{\rho=0}\,\,.
\label{eq:50}
\eeq
Here 
$P_{q}^{g}(x)=C_{F}[1+(1-x)^{2}]/x$ is the usual splitting
function for $q\to g q$ process,
$
M=Ex(1-x)\,
$
is the reduced "Schr\"odinger mass",
$F$ is the solution to the radial Schr\"odinger 
equation for the azimuthal quantum number $m=1$ 
\beq
\hspace{-.2cm} i\frac{\partial F(\xi,\rho)}{\partial \xi}=
\left[-\frac{1}{2M}\left(\frac{\partial}{\partial \rho}\right)^{2}
+v(\rho,x,z-\xi)
+\frac{4m^{2}-1}{8M\rho^{2}}
+\frac{1}{L_{f}}
\right]F(\xi,\rho)\,
\label{eq:60}
\eeq
with the boundary condition
$F(\xi=0,\rho)=\sqrt{\rho}\sigma_{3}(\rho,x,z)
\epsilon K_{1}(\epsilon \rho)$  
($K_{1}$ is the Bessel function),
$L_{f}=2M/\epsilon^{2}$
with $\epsilon^{2}=m_{q}^{2}x^{2}+m_{g}^{2}(1-x)^{2}$,
$\sigma_{3}(\rho,x,z)$ is the cross section of interaction
of the $q\bar{q}g$ system with a medium constituent
located at $z$.
The potential $v$ in (\ref{eq:60}) reads
\beq
v(\rho,x,z)=-i\frac{n(z)\sigma_{3}(\rho,x,z)}{2}\,.
\label{eq:70}
\eeq
The $\sigma_{3}$ is given by
\cite{NZ_sigma3}
\beq
\sigma_{3}(\rho,x,z)=\frac{9}{8}
[\sigma_{q\bar{q}}(\rho,z)+
\sigma_{q\bar{q}}((1-x)\rho,z)]-
\frac{1}{8}\sigma_{q\bar{q}}(x\rho,z)\,,
\label{eq:80}
\eeq
where
\beq
\sigma_{q\bar{q}}(\rho,z)=C_{T}C_{F}\int d\qb
\alpha_{s}^{2}
\frac{[1-\exp(i\qb\ro)]}{[q^{2}+\mu^{2}_{D}(z)]^{2}}\,
\label{eq:90}
\eeq
is the local  dipole cross section for the color singlet $q\bar{q}$ pair,
$C_{F,T}$ are the color Casimir for the quark and thermal parton 
(quark or gluon), $\mu_{D}$ is the local Debye mass.
\begin{figure}[!h] 
\begin{center}
\includegraphics[height=1cm]{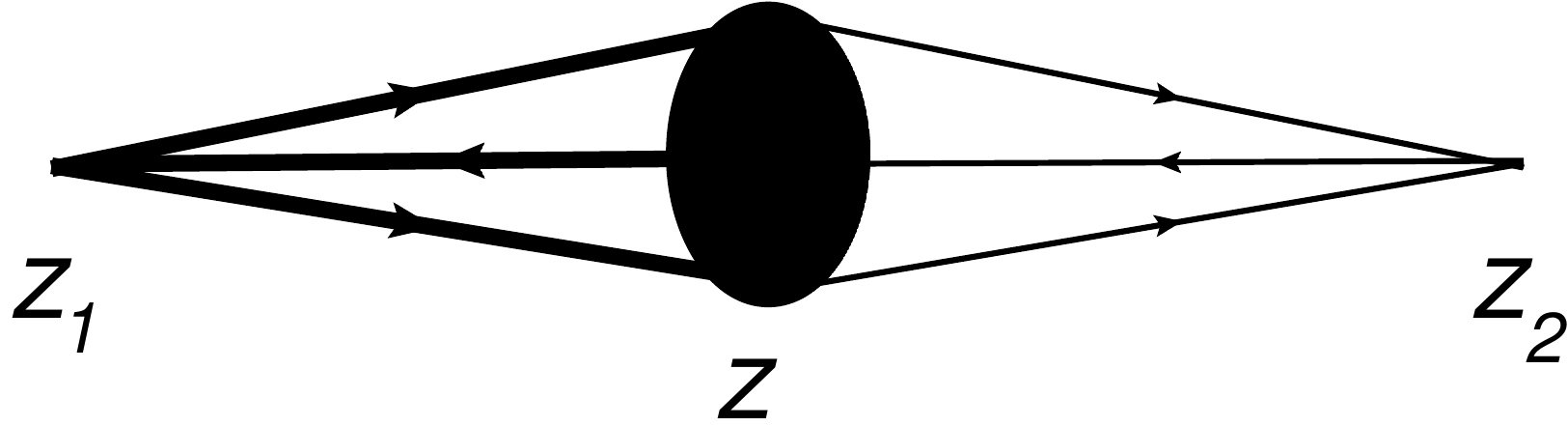}  
\end{center}
\caption[.]{Diagrammatic representation of the effective Bethe-Heitler
  cross section for $a\to bc$ process in terms of the dressed (left)
  and bare (right)
  Green functions describing $z$-evolution of the $\bar{a}bc$ system.
  The central blob corresponds to the potential $v$.
  Integration over $z_1$ (from $0$ to $z$) and $z_2$ (from $z$ to $\infty$)
  is implied.}
\end{figure}

Diagrammatically, the effective Bethe-Heitler cross section (\ref{eq:50}),
for any partonic process $a\to bc$,
is given by the graph shown in Fig.~1 where the left
and right parts correspond to the dressed and bare Green functions
describing $z$-evolution of the $\bar{a}bc$ three-body system
(i.e. $\bar{q}qg$ for $q\to qg$ process).
The central black blob in Figs.~1 describes
interaction of the three-body state with a medium constituent.
For RHIC and LHC conditions the dominating contribution to
the effective Bethe-Heitler cross section comes from $N=1$ scattering.
This means that the dressed Green function in Fig.~1 is close to the bare one.
Therefore in this regime the average $z-z_1$ is close to $z_2-z$.
For generalization of the above formulas to the case of
running $T$-dependent coupling
one should modify $\alpha_s$ that appears
on the right hand side of (\ref{eq:50}), which
comes from product
of the QCD couplings in the decay vertices at $z_1$ and $z_2$ in Fig.~1,
and $\alpha_s^2$
in formula for the dipole cross section (\ref{eq:90}).
In the latter case it is natural simply to replace the fixed $\alpha_s$
by the local running coupling $\alpha_s(q,T(z))$.
However, the situation is more complicated
for $\alpha_s$ on the right hand side of (\ref{eq:50}).
In terms of the variable $\xi$ in (\ref{eq:50}) $z_1=z-\xi$.
As we said above, for the dominating $N=1$ scattering term
on the average $z-z_1\sim z_2-z$.  
We will use this approximation
for the whole
effective Bethe-Heitler cross section.
Then, for the temperatures at the decay vertices one can take
$T(z\pm \xi)$. This approximation should be reasonable due to
a smooth dependence of $T$ on the proper time ($T\propto \tau^{-1/3}$) and
a smooth (logarithmic) dependence of $\alpha_s$ on
the QGP temperature.  
Since we work in the coordinate space the virtualities at the decay vertices
do not appear in our formulas.
Qualitatively, from the
uncertainty relation, one can obtain that $Q^2\sim 1/\rho^2$, where
$\rho$ is the transverse size of the three-body state at $z$.
Similarly to our previous analyses
of jet quenching with a unique $T$-independent running $\alpha_s$,
we determine the virtuality for these vertices as $Q^2(\xi)=a/\xi$
with $a=1.85$ \cite{Z_coll}. This parametrization accounts for the
Schr\"odinger diffusion relation, that gives $\rho^2\sim \xi/M$,
and the value of the parameter $a$ has been adjusted to reproduce the $N=1$
scattering contribution evaluated in the ordinary Feynman diagrammatic
approach \cite{Z_kin}. Thus, for calculations
with the $T$-dependent running coupling we replace 
the fixed $\alpha_s$ on the
right-hand side of (\ref{eq:50}) by
$\sqrt{\alpha_s(Q(\xi),T(z-\xi))\alpha_s(Q(\xi),T(z+\xi))}$.
We checked that the version with $T(z\pm \xi)$ replaced by $T(z)$
gives practically the
same results, i.e. the effect of the finite separation
between the decay vertices is small. This occurs because the dominating
contribution to the radiative energy loss comes from
gluons with the formation length which is considerably smaller
than the QGP size.

First principle calculations of the $\alpha_s(Q,T)$ in the QGP are not yet
available.
In the lattice analysis \cite{Bazavov_al1},
via calculation of the free energy of a static heavy
quark-antiquark pair, there have been obtained an effective in-medium coupling
$\alpha_s(r,T)$ in the coordinate representation.
The results of \cite{Bazavov_al1} show that $\alpha_s(r,T)$ at $r\ll 1/T$
 becomes close to the ordinary vacuum QCD coupling
$\alpha_s(Q)$ with $Q\sim 1/r$. In the infrared region
$\alpha_s(r,T)$ reaches maximum at
$r\sim 1/\kappa T$ with $\kappa\sim 4$ and then with increasing $r$ it falls to
zero. With identification $r\sim 1/Q$, this pattern is qualitatively
similar to that obtained for $\alpha_{s}(Q,T)$ in the momentum representation
within the functional renormalization group calculations \cite{RG1}. 
Motivated by the results of \cite{Bazavov_al1,RG1},
we use parametrization of $\alpha_s(Q,T)$ in the form
\beq
\alpha_s(Q,T) = \begin{cases}
\dfrac{4\pi}{9\log(Q^2/\Lambda_{QCD}^2)}  & \mbox{if } Q > Q_{fr}(T)\;,\\
\alpha_{s}^{fr}(T) & \mbox{if }  Q_{fr}(T)\ge Q \ge cQ_{fr}(T)\;, \\
\alpha_{s}^{fr}(T)\times(Q/cQ_{fr}(T)) & \mbox{if }  Q < cQ_{fr}(T)\;, \\
\end{cases}
\label{eq:100}
\eeq
where
$Q_{fr}=\Lambda_{QCD}\exp\left\lbrace
{2\pi}/{9\alpha_{s}^{fr}}\right\rbrace$ (in the present analysis we
take $\Lambda_{QCD}=200$ MeV), and $c<1$.
The parameter $c$ defines the width of the plateau
where $\alpha_s$ equals its maximum value $\alpha_s^{fr}$.
For our basic version we take $c=0.8$. We have also performed
calculations for $c=0$.
The case $c=0$ is similar to the model with
a frozen QCD coupling in the infrared region at $T=0$ \cite{Stevenson,DKT}.
In \cite{DKT} it was called the $F$-model.
For $c\sim 1$ the parametrization (\ref{eq:100})
is qualitatively similar
to the $G$-model of the vacuum $\alpha_s$ of \cite{DKT}. 
We take $Q_{fr}=\kappa T$, and perform fit of the free parameter $\kappa$ using
the data on the nuclear modification factor $R_{AA}$.
From the lattice results of \cite{Bazavov_al1} one can expect
that $\kappa\sim 4$. But since
the relation $r\sim 1/Q$ is of qualitative nature,
our parameter $\kappa $ may differ from that in the lattice calculations
in the coordinate space. Also, one should bear in mind that
in the infrared region the effective coupling becomes process dependent
\cite{Bazavov_al1}.
We also present the results
for a unique $\alpha_s$ in the whole QGP with the $T$-independent free
parameter $\alpha_s^{fr}$.

{\bf 3. Numerical results}.
We perform calculations for the QGP fireball
with purely longitudinal Bjorken's 1+1D expansion \cite{Bjorken}, 
which
gives
proper time dependence of the entropy density
$s(\tau)/s(\tau_0)=\tau_0/\tau$, where $\tau_0$ is the QGP thermalization
time.
We take $\tau_{0}=0.5$ fm.
As in \cite{RAA20}  we take a linear $\tau$-dependence
$s(\tau)=s(\tau_0)\tau/\tau_0$ for $\tau<\tau_{0}$.
We neglect variation of the initial QGP density with the 
transverse coordinates across the overlapping area of two colliding
nuclei. The average initial temperature has been evaluated in the Glauber
wounded nucleon model \cite{KN-Glauber} (with
parameters from \cite{Z_MC1,Z_MC2}). For central collisions this gives
$T_0\sim 320(420)$ MeV for 0.2 TeV Au+Au(5.02 TeV Pb+Pb) collisions.  
We refer the reader to \cite{RAA20} for more details
on the model and parameters of the QGP fireball.
For the $T$-dependence of $\alpha_s$ and the Debye mass
we use the temperature extracted from the entropy $s(\tau)$ using the 
lattice entropy density obtained in \cite{t-lat}. For a given entropy,
this temperature is somewhat larger than the ideal gas temperature.
As in \cite{RAA20}, we use the Debye mass obtained in
the lattice analysis \cite{Bielefeld_Md}, 
and take $m_{q}=300$ and $m_{g}=400$ MeV 
for the light quark and gluon quasiparticle masses
in the QGP \cite{LH}.

\begin{table}
  \hspace{-2cm}
         {
  \begin{tabular}{c|cc|cc}
    \hline
& \multicolumn{2}{c}{$\alpha_s(Q,T)$} &
\multicolumn{2}{|c}{$\alpha_s(Q)$} \\
\cline{2-5}
& $\kappa $ & $\chi^2/d.p.$ &$\alpha_s^{fr}$ &  $\chi^2/d.p.$  \\
\hline
PHENIX Au+Au~0.2~TeV  &$2.65$ &$0.167(0.71,0.81)$
&$0.698$ & $0.157(4.4,4.75)$\\
\hline
ALICE Pb+Pb~5.02~TeV  &$3.19$
& $0.46(0.68)$ &
$0.464(0.464)$ & $0.56(0.88)$\\
\hline
ATLAS Pb+Pb~5.02~TeV   &$3.48(3.46)$ &
$0.37(0.22)$ &$0.439(0.439)$ & $0.33(0.2)$\\
\hline
CMS Pb+Pb~5.02~TeV   &$3.99(3.81)$ & $0.58(0.25)$ &
$0.403(0.412)$ & $0.46(0.21)$\\
\hline
All LHC Pb+Pb~5.02~TeV   &$3.33(3.28)$ & $1.04(0.96)$ &
$0.451(0.455)$ & $0.93(0.96)$\\
\hline
  \end{tabular}
}  
  \caption{
    Optimal values of $\kappa$ and $\alpha_s^{fr}$ (for $c=0.8$ in formula
    (\ref{eq:100})) 
    and corresponding $\chi^2/d.p.$
    for different data sets.
    For LHC the results are presented for
    fits for the data points with
    $9<p_T<120$ and $9<p_T<22$ GeV (the numbers in brackets).
    For RHIC the fits are performed
    for the data points with $p_T>9$ GeV.
    The numbers in brackets for
    $\chi^2/d.p.$ for RHIC give $\chi^2/d.p.$ obtained with the LHC optimal
    parameters $\kappa/\alpha_s^{fr}$ obtained for the LHC fits
    for $9<p_T<22$ GeV and $9<p_T<120$ GeV.
\label{chi2fit}}
\end{table}
\begin{figure}[!h]
\begin{center}
\includegraphics[height=8.5cm]{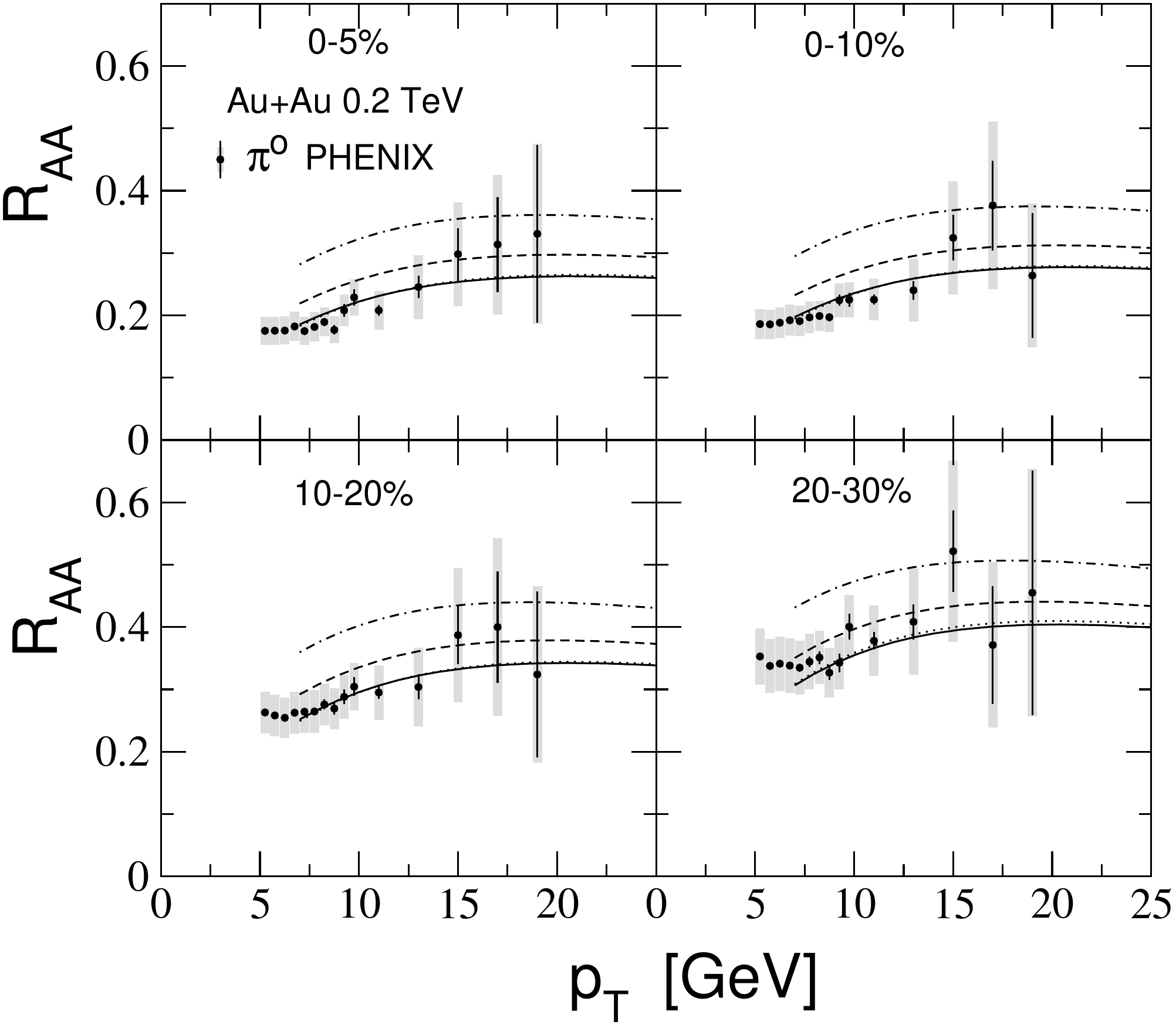}  
\end{center}
\caption[.]{$R_{AA}$ of $\pi^{0}$ for $0.2$ TeV  Au+Au collisions
  for different centrality bins
  from our calculations for $T$-dependent $\alpha_s$ (solid and dashed) and
  $T$-independent $\alpha_s$ (dotted and dash-dotted)
  compared to data from PHENIX \cite{PHENIX_r}.
  The solid and dotted curves are for $\kappa=2.65$ and
  $\alpha_s^{fr}=0.698$ obtained by fitting the PHENIX $R_{AA}$ data set
  for $p_T>9$ GeV.
 The dashed and dash-dotted lines are for $\kappa=3.28$ and
 $\alpha_s^{fr}=0.455$ obtained by fitting the LHC  $R_{AA}$ data sets
 for $9<p_T<22$ GeV.}
\end{figure}
\begin{figure} [!h]
\begin{center}
\includegraphics[height=8.5cm]{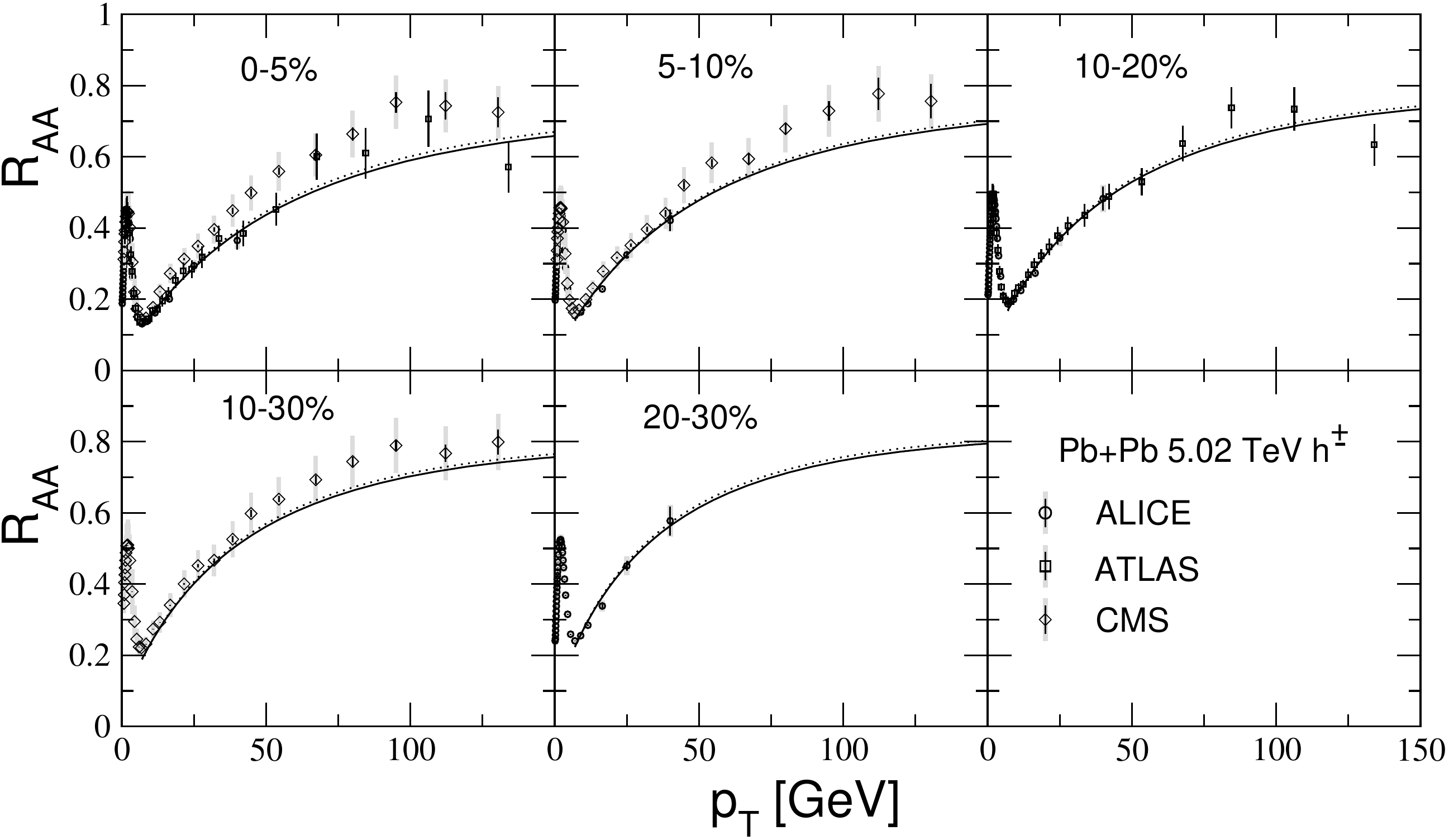}  
\end{center}
\caption[.]
        {
          $R_{AA}$ of charged hadrons for $5.02$ TeV Pb+Pb collisions
          for different centrality bins.
          Solid: calculations for $T$-dependent $\alpha_s$
          with $\kappa\approx 3.33$.
Dotted: calculations for $T$-independent $\alpha_s$
with $\alpha_s^{fr}\approx0.451$. $\kappa$ and $\alpha_s^{fr}$ are
obtained by fitting $R_{AA}$ in the range
$9<p_T<120$ GeV.
  Data points are from ALICE \cite{ALICE_r502}, ATLAS \cite{ATLAS_r502},
  and CMS \cite{CMS_r502}.
}
\end{figure}

We have performed the $\chi^2$ fit of the free parameters
$\kappa$/$\alpha_s^{fr}$ using the
the RHIC and LHC data on $R_{AA}$ for centralities smaller than 30\%.
We use the data 
for $0.2$ TeV Au+Au collisions  at RHIC from
PHENIX \cite{PHENIX_r} for $\pi^0$-meson   and
for $5.02$ TeV collisions Pb+Pb at the
LHC
from ALICE \cite{ALICE_r502}, ATLAS \cite{ATLAS_r502},
and CMS \cite{CMS_r502} for charged hadrons.
We take
$p_{T,min}=9$ GeV\footnote{The $\chi^2$ fits with
  $p_{T, min}\sim 7-10$ GeV give very similar results.
However, the inclusion of the data points
 with $p_T\lsim 7-8$ GeV does not make sense, since at such $p_T$
 the non-fragmentation recombination
 mechanism \cite{Fries,Greco}
 may become important. It is quite likely that just this mechanism
 causes the growth of $R_{AA}$ at $p_T\lsim 6-7$ GeV for the LHC energies.}.
For the PHENIX data \cite{PHENIX_r}
we include all data points with $p_T>p_{T,min}$.
For the LHC data we perform fitting for $p_{T,max}=120$ and $22$ GeV.
The latter value
seems to be preferable for studying
the variation of $\alpha_s$ from RHIC to LHC, since for the PHENIX
data $p_T<20$ GeV. But we have found that the LHC fits
for $p_{T,max}=120$ and $22$ GeV give very similar results.
We calculate $\chi^2$ as
\beq
\chi^2 = \sum_i^N \frac{(f_i^{exp} - f_i^{th})^2}
    {\sigma_{i}^2}\,,
    \label{eq:130}
    \eeq
where $N$ is the number of the data points, the squared errors
include the systematic and statistic errors
$\sigma_i^2=\sigma_{i,stat}^2+\sigma_{i,sys}^2$.
In calculating $\chi^2$ as functions of free parameters $\kappa$
and $\alpha_s^{fr}$ we have used 
the theoretical $R_{AA}$ obtained with the help of a cubic spline
interpolation from the grids calculated with steps
(in $\kappa$ and $\alpha_s^{fr}$)
$\Delta \kappa/\kappa,\Delta \alpha_s^{fr}/\alpha_s^{fr}\sim 0.05$.
The optimal values of $\kappa$ and $\alpha_s^{fr}$ with corresponding
values of $\chi^2/d.p.$ ($\chi^2$ per data point)  are summarized in Table I.
In Table I we show the results for RHIC and LHC separately.
We also performed fitting for the combined RHIC plus LHC data set
(not shown). In this case the results are very close to that
for the LHC data set alone. This occurs because the number of the
data points for the LHC data set is much bigger than for the RHIC data
set.
From Table I one can see that the LHC data give somewhat bigger
value of the optimal parameter $\kappa$. However, the difference
is not very big.
To illustrate better the difference between RHIC and LHC, in Table I for the
PHENIX data set besides $\chi^2/d.p.$ for
$\kappa$ and $\alpha_s^{fr}$
fitted to the PHENIX data we also give
$\chi^2/d.p.$ for the optimal
$\kappa$ and $\alpha_s^{fr}$ obtained for the LHC data set.
One case see that for the $T$-dependent $\alpha_s$
the LHC value of $\kappa$
gives rather good fit quality $\chi^2/d.p.\approx 0.7-0.8$, while
for the $T$-independent $\alpha_s$
for the LHC optimal parameter $\alpha_s^{fr}$ we
have $\chi^2/d.p.\approx 4.4-4.8$,
that says about a rather strong disagreement with the PHENIX data.

In Figs.~2--3 we compare our results for $R_{AA}$
with the RHIC data from PHENIX for $\pi^{0}$-meson in  
$0.2$ TeV Au+Au collisions \cite{PHENIX_r} and
the LHC data \cite{ALICE_r502,ATLAS_r502,CMS_r502}
for $h^{\pm}$ in $5.02$ TeV Pb+Pb collisions.
One can see that for the optimal parameters (separately for RHIC and LHC)
agreement with the data is quite good for both the versions.
However, the situation with a joint description of the RHIC
and LHC data is very different for the $T$-dependent and $T$-independent
couplings.
To visualize better this difference in Fig.~2, in addition to
predictions for $R_{AA}$ in Au+Au collisions obtained with
the optimal parameters
fitted to the PHENIX data, we also plot the results for
the optimal parameters fitted to the LHC data.
As one can see, for the version with $T$-dependent coupling
the LHC value of $\kappa$ leads to not bad agreement with the PHENIX data.
While for the version with $T$-independent coupling the curves
for the LHC value of $\alpha_s^{fr}$ overshoot the data considerably
at $p_T\lsim 15$ GeV. For the $T$-dependent version
some  overshooting at $p_T\lsim 15$ GeV also exists, but it is rather small.
We conclude from comparison with experimental data shown in Figs.~2--3
and results of our fits given in Table I, that the
$T$-dependence of the QCD coupling may strongly reduce the difference
between the optimal $\alpha_s$ for RHIC and LHC\footnote{Note that
  for the LHC data on 2.76 TeV Pb+Pb collisions
\cite{ALICE_r276,ATLAS_r276,CMS_r276}
  the situation is the same. Our
  calculations show that the optimal parameters (and quality of the fits)
in this case are very close to that for 5.02 TeV Pb+Pb collisions.}.
We have checked that, in principle, for the $T$-dependent coupling
by a relatively small increase of $\alpha_s(Q,T)$
at $Q\sim (1-3)\Lambda_{QCD}$, as compared to the one-loop formula
used in (\ref{eq:100}), for $\kappa$ fitted to the LHC data
one can significantly improve agreement
with the RHIC data in the low $p_T$ region.
Such an increase of the $\alpha_s(Q,T)$ is not unrealistic, e.g., it
may mimic an enhancement of the induced gluon emission at $T\sim T_c$
\cite{Liao_magQGP,Z_magQGP}
in the presence color-magnetic monopoles \cite{magQGP1}.

We have also calculated the azimuthal asymmetry $v_2$.
Although we have fitted the optimal parameters to the data on $R_{AA}$,
for both the $T$-dependent
and $T$-independent versions we have obtained a quite reasonable
agreement with the $v_2$
data as well. For the $T$-dependent version, we obtained a bit
bigger $v_2$. This occurs due to some enhancement of the contribution
to the energy loss from the later, low temperature, stage of the
QGP evolution, which has a bigger initial fireball azimuthal asymmetry.

The above results have been obtained for parametrization (\ref{eq:100}) with
$c=0.8$.
The results from calculation for $c=0$ in (\ref{eq:100}), i.e. for
flat $\alpha_s$ at $Q<Q_{fr}$, turn out to be very similar to that
for $c=0.8$. We obtained rather good agreement with the RHIC and
LHC data for the versions with $T$-dependent and $T$-independent coupling.
However, similarly to the case $c=0.8$, the latter version leads to a
considerable disagreement between free
parameters for RHIC and LHC. While the former version
largely eliminates this disagreement.

{\bf 4. Summary}.
We have studied the influence of the temperature dependence
of running coupling on the variation of jet quenching from the RHIC
to LHC energies within the LCPI \cite{LCPI1} approach to the induced
gluon emission. 
The calculations are performed using the method suggested
in \cite{RAA04,RAA08}.
For our basic version we use parametrization of running
coupling $\alpha_s(Q,T)$
which has a short plateau $\alpha_{s}^{fr}$ around $Q_{fr}\sim \kappa T$, and
then falls $\propto Q$ at small $Q$.
This ansatz is motivated by the lattice calculation of the effective
QCD coupling in the QGP \cite{Bazavov_al1} and the results
obtained within the functional renormalization group \cite{RG1}.
We have determined the optimal values of the parameter $\kappa$ fitting
the data on the nuclear modification factor $R_{AA}$ in $0.2$ TeV
Au+Au collisions at RHIC and in $5.02$ TeV Pb+Pb collisions at the LHC.
We have found that the RHIC data require somewhat smaller value
of the parameter $\kappa$ than the LHC data. But nevertheless
the theoretical $R_{AA}$ for $0.2$ Au+Au collisions
calculated with the optimal $\kappa$ adjusted to fit the LHC data,
is in reasonable agreement with the RHIC data
($\chi^2/d.p.\approx 0.7-0.8$).
This differs drastically from
the results for the $T$-independent
$\alpha_s^{fr}$, which leads to rather strong disagreement with the RHIC
data ($\chi^2/d.p.\approx 4.4-4.8$)  for the optimal value $\alpha_s^{fr}$ fitted to the LHC data.
Thus, our analysis shows that
the $T$-dependent $\alpha_s$ may largely eliminate the problem of different
optimal QCD coupling for the RHIC and LHC energies. 
For parametrization with flat $\alpha_s$ at $Q<Q_{fr}$ with $Q_{fr}=\kappa T$
we obtained very similar results. 
Our results may be viewed as the first direct evidence of the increase
of the thermal suppression
of $\alpha_s$ with rising QGP temperature.

\begin{acknowledgments}
  This work was performed under the Russian Science Foundation grant 20-12-00200
at Steklov Mathematical Institute.
\end{acknowledgments}

\end{document}